\documentclass{article}

\usepackage{arxiv}

\usepackage{comment}
\usepackage{lipsum} 
\usepackage{cite}
\usepackage{amsmath,amssymb,amsfonts}
\usepackage{algorithmic}
\usepackage{graphicx}

\usepackage{graphicx}
\usepackage{textcomp}
\usepackage{xcolor}
\def\BibTeX{{\rm B\kern-.05em{\sc i\kern-.025em b}\kern-.08em
    T\kern-.1667em\lower.7ex\hbox{E}\kern-.125emX}}
\usepackage{amsthm}                              
\usepackage{mathtools}
\mathtoolsset{showonlyrefs}                         
\usepackage{amsfonts,amssymb,amsmath}
\usepackage[T1]{fontenc}
\usepackage{xcolor}
\usepackage{pifont}

\usepackage{graphicx}
\usepackage[normalem]{ulem}

\newenvironment{pro}{\begin{proof}}{\end{proof}}
\usepackage{url}

\usepackage{subcaption}
\usepackage{algorithm2e}
\usepackage{algorithmic}
\usepackage[compact]{titlesec}
\usepackage{chngcntr}
 \titlespacing{\section}{0pt}{1ex}{0ex}
 \titlespacing{\subsection}{0pt}{0ex}{0ex}
 \titlespacing{\subsubsection}{0pt}{0ex}{0ex}
 \titlespacing{\paragraph}{0pt}{0ex}{0ex}
\usepackage{hyperref}
\usepackage{cleveref}
\newtheorem{lemma}{Lemma}
\newtheorem{theorem}{Theorem}
\newtheorem{claim}{Claim}

\usepackage{enumitem}
\usepackage{amsfonts,amsmath,amssymb}
\renewcommand{\paragraph}[1]{\smallskip\noindent\textbf{#1.}}

\title{ScaloWork: Useful Proof-of-Work with Distributed Pool Mining}

\author{
 Diptendu Chatterjee \\
  Department of Computer Science and Information Systems\\
  BITS Pilani, KK Birla Goa Campus, India\\
  \href{mailto:diptenduc@goa.bits-pilani.ac.in}{diptenduc@goa.bits-pilani.ac.in} \\
   \And
 Avishek Majumder \\
  School of Coumputer Science\\
  UPES, Dehradun, India\\
    \href{mailto:avishek.majumder1991@gmail.com}{avishek.majumder1991@gmail.com}\\
  \And
 Subhra Mazumdar \\
  Department of Computer Science and Engineering\\
  IIT Indore, India\\
  \href{mailto:subhra.mazumdar@iiti.ac.in}{subhra.mazumdar@iiti.ac.in}
}

\begin{document}
\maketitle
\begin{abstract}
Bitcoin blockchain uses hash-based Proof-of-Work (PoW) that prevents unwanted participants from hogging the network resources. Anyone entering the mining game has to prove that they have expended a specific amount of computational power. However, the most popular Bitcoin blockchain consumes 175.87 TWh of electrical energy annually, and most of this energy is wasted on hash calculations, which serve no additional purpose. Several studies have explored re-purposing the wasted energy by replacing the hash function with meaningful computational problems that have practical applications. Minimum Dominating Set (MDS) in networks has numerous real-life applications. Building on this concept, {\em Chrisimos} [TrustCom '23] was proposed to replace hash-based PoW with the computation of a dominating set on real-life graph instances. However, \emph{Chrisimos} has several drawbacks regarding efficiency and solution quality. 

This work presents a new framework for Useful PoW, {\em ScaloWork}, that decides the block proposer for the Bitcoin blockchain based on the solution for the dominating set problem. \emph{ScaloWork} relies on the property of graph isomorphism and guarantees solution extractability. We also propose a distributed approach for calculating the dominating set, allowing miners to collaborate in a pool. This enables {\em ScaloWork} to handle larger graphs relevant to real-life applications, thereby enhancing scalability. Our framework also eliminates the problem of free-riders, ensuring fairness in the distribution of block rewards.  We perform a detailed security analysis of our framework and prove our scheme as secure as hash-based PoW. We implement a prototype of our framework, and the results show that our system outperforms \emph{Chrisimos} in all aspects.
\end{abstract}

\section{Introduction}\label{sec:introduction}
Adam Back’s Hashcash \cite{back2002hashcash} forms the basis for hash-based PoW consensus protocols in the Bitcoin network. 
Each block added to the chain has a transaction set and a block header. 
To add a block, a miner has to find a {\em nonce value}, such that the hash of her block header, along with the nonce, satisfies a predefined (difficult) target \cite{antonopoulos2014mastering}.
Finding the appropriate nonce for the given block is a computationally intensive task. 
As per the report in December 2024 \cite{energy}, Bitcoin is estimated to have an annual power consumption of 175.87 \texttt{terawatt-hours (TWh)} - more than the power consumption of many countries, including countries like Norway. 
Most of this power is consumed to calculate the hash to mine a block, which has no further utility.

Many cryptocurrencies are switching to alternate consensus protocols to overcome
this problem. 
However, alternate consensus has its own problems. 
For example, Proof-of-Stake (PoS) \cite{king2012ppcoin,kiayias2017ouroboros} uses a staking system where a certain amount of capital in the form of the network’s tokens is required to become a \textcolor{black}{validator}.
Proof-of-Authority \cite{poa} eliminates democracy and moves the power to the {\em ``richest in the room''}. 
Proof-of-Space \cite{dziembowski2015proofs,park2018spacemint} needs more storage space when more miners are added to the network. 
Proof-of-Spacetime \cite{pos-dis} requires users to lock up a certain amount of coins to store data on the Blockchain, which can create an entry-level barrier for new users. 
Proof-of-Burn \cite{pob-ds} wastes coins, as mining power is proportional to the amount of money a participant is willing to burn. 

The security is meant to be derived directly from the perceived economic value of the network or how expensive it is to purchase a majority stake. While it is true that the second most popular consensus protocol than PoW, PoS verification, is less energy-intensive than the PoW system currently in place, the major concern with PoS networks is the level of centralization and its subsequent impact on the security of the network \cite{switchBitcoin}. On the other hand, hash-based PoW is democratic, where a newcomer and a seasoned veteran have the same voice and power.


     Any computational puzzle requiring substantial resources to solve is compatible with Nakamoto's Proof-of-Work consensus mechanism \cite{nakamoto2008bitcoin}. One can replace the conventional nonce-finding hash-based protocol with a task that has either a commercial purpose or an academic utility. The requirement led to the concept of {\em ``useful Proof-of-Work''} or {\em ``Proof-of-Useful-Work''} (PoUW). The term was first mentioned in \cite{becker2013can} and later formalized by Ball et al. \cite{ball2017proofs}. Further constructions for PoUW mining were given by Loe et al. \cite{loe2018conquering} and Dotan et al. \cite{dotan2020proofs}. In all these previous approaches, \textcolor{black}{the system's security was not rigorously analyzed} and in many cases, the attacker can manipulate the graph instances provided to the system and rig the mining game. A hybrid approach to mining \cite{philippopoulos2020difficulty}  combines hash value calculations with difficulty-based incentives for problem-solving. However, miners can be dishonest, find multiple solutions to a real-life instance, mine a longer chain using all the solutions, and perform chain reorganization.
\textcolor{black}{Zheng et al. introduced \emph{AxeChain} \cite{zheng2020axechain} that uses the computing power of blockchain to solve NP-complete problems. However, the protocol is susceptible to solution stealing, as all the miners retrieve the highest priority problem from the problem queue. Since all the miners work on the same problem, lazy miners can steal solutions from their neighbors and broadcast the best solution to claim rewards}. Another PoUW protocol \emph{Ofelimos} \cite{fitzi2022ofelimos} is based on the doubly parallel local search (DPLS) algorithm. DPLS represents a general-purpose stochastic local-search algorithm with an exploration algorithm component. However, their approach does not consider the difficulty of NP-hard problem instances, which might lead to unfairness during block generation. A recent work called Combinatorial Optimization Consensus Protocol (COCP) \cite{todorovic2022proof} proposes efficient utilization of computing resources by providing valid solutions for the real-life instances of any combinatorial optimization problem. A major drawback of the scheme is that the miner, upon finding the solution to a combinatorial optimization problem, sends it to a solution pool controlled by a centralized entity. The pool returns the best solution.

\textcolor{black}{Many real-world systems can be modeled as graphs, e.g., communication networks, transportation systems, IoT networks, computer communication networks, interconnection networks, etc. \cite{sym12111885}}. Solving graph-theoretic problems like the dominating set problem can directly contribute to advancements in these fields. \textcolor{black}{We observe that NP-complete graph theoretic problems like finding minimum dominating set, minimum graph coloring, etc., have practical utility. 
This class of problems perfectly fits into PoW-based blockchain systems because identifying the solution in the first place requires no known polynomial algorithm (thus, it is ``hard'' to compute). Still, the solution to the problem can be
verified in polynomial time \cite{oliver2017proposal}.} While other graph-theoretic problems may also be viable, the dominating set problem stands out as it directly applies to real-world scenarios, such as network optimization, IoT device coverage, and resource allocation. \textcolor{black}{To effectively influence network participants, a critical feature of networks is rapid and efficient communication. Minimum Dominating Set (MDS) plays a pivotal role in this process - by identifying the smallest subset of nodes that can reach all others in one step, MDS ensures optimal information propagation, reduces latency, and minimizes resource usage. This structural efficiency is vital for maintaining real-time coordination, viral content spread, and resilience against disruptions in decentralized systems like blockchain or peer-to-peer networks.} Online social network sites like Facebook, LinkedIn, and X rely on dominating sets for large networks to realize the desired goal and spread ideas and information within a group. It can be used for targeted advertisements and alleviate social problems \cite{wang2014domination}. Dominating sets play an important role in controlling the spreading of rumours and fake information \cite{alipour2020distributed}.

\textcolor{black}{The dominating set problem works on any graph, allowing endless diversity in problem instances. In applications like viral marketing, emergency alerts, or sensor networks, minimizing the number of active nodes (e.g., influencers, relay stations) reduces cost and energy consumption. Graphs taken from real-world datasets or structured to suit specific use cases ensure miners cannot precompute solutions or reuse the same solution across different blocks, maintaining the integrity of the PoW mechanism. Several real-world networks need dynamic or periodic MDS updates due to structural changes. Power grid require phasor measurement units (PMUs) to monitor voltage, current, and frequency in real-time \cite{haynes2002domination}. When a failure (e.g., line fault, generator outage) occurs, the grid must quickly detect and isolate the issue. In Vehicular ad-hoc networks (VANETs), to select the best RSU (Roadside Unit), MDS plays a role in establishing connection between the Consumer vehicle and smartCloud vehicle for the access of various services \cite{chinnasamy2019minimum}.  As long as networks exist—and they always will—researchers and engineers will need better ways to compute, maintain, and optimize dominating sets.}

A recent work, \emph{Chrisimos} \cite{chatterjee2023chrisimos}, incorporated the problem of finding a dominating set of real-life graph instances as a replacement for finding nonce in Hash-based PoW. In this protocol, a utility company announces a graph instance to the Blockchain network and the reward for solving the dominating set problem. 
Any miner that finds a dominating with the lowest cardinality within the given epoch wins the mining game, becomes the block proposer, and earns the block reward. 
Though the idea is novel, we identified several shortcomings. 
If a miner returns the dominating set on the original graph, the chance is high that other miners will steal and broadcast it as their own solution. 
To ensure that lazy malicious miners do not steal other honest miners' solutions, \emph{Chrisimos} probabilistically extends the graph instance to twice the size. This puts a lot of computation and storage overhead on the miners mining a block. Additionally, \emph{Chrisimos} suffers from the problem of solution extractability. Finding a good dominating set on the extended graph does not guarantee that one can extract a good solution for the original graph. Also, the probabilistic extension rule is not fair, as one miner might have the advantage of topology over the other while finding the solution based on the extension strategy applied to the original instance. Another drawback of \emph{Chrisimos} is that it cannot process large graphs due to the probabilistic extension. This we justify in detail in Section~\ref{performace}. 
\textcolor{black}{The framework of \emph{Chrisimos} did not also consider pool mining by distributing the task.}
\begin{table*}[!htb]
    \centering
    \resizebox{\textwidth}{!}{
    \begin{tabular}{|c|c|c|c|c|c|c|c|c|}
    \hline
          & Usefulness & Block proposal  & Verification Soundness &Fairness of & Adjustable &Solution  &Scalability &Security\\
             & &efficiency &\& Efficiency &Block Reward &Hardness &Extractability & &\\
    \hline
    Primecoin \cite{king2013primecoin} &Limited &Yes &With high probability &Yes &Yes &Yes &No &Yes\\
    Dotan et al. \cite{dotan2020proofs} &Yes &Yes &Yes &Yes &Yes &Yes &No &No\\
    DLchain \cite{tian2020dlchain} &Yes &Yes &Fails upon fork &Yes &No &Yes &Yes &Not clear\\
    Coin.AI \cite{baldominos2019coin} &Yes &Not clear &Limited &Not clear &No &Yes &No &Not clear\\
    AxeChain \cite{zheng2020axechain} &Yes &Not clear &Not clear &Yes &Yes &Yes &No &No\\
    
        Ofelimos \cite{fitzi2022ofelimos} &Yes &Yes &Yes &Not clear &No &Yes &Yes &Yes\\
        COCP \cite{todorovic2022proof} &Yes &Yes &Yes &No &Not clear &Yes &No &Not clear\\
        Chrisimos \cite{chatterjee2023chrisimos} &Yes &No &Yes &No &Yes &No &No &Yes\\
        ScaloWork (Our Work) &Yes &Yes &Yes &Yes &Yes &Yes &Yes &Yes\\
    \hline
    \end{tabular}
    }
    \caption{Comparative Analysis of \emph{ScaloWork} with state-of-the-art}
    \label{tab:my_label}
\end{table*}

So the following questions still remained unanswered for a PoUW-based consensus algorithm:
\begin{quote}
    {\em ``Can we devise a framework for useful proof of work with the following properties? -
        (a) tasks must have social and economic utility,
        (b) no additional preprocessing on the task,
        (c) the verifier must accept only the correct solution,
        (d) the verification must be efficient,
        (e) the framework must guarantee solution extractability,
        (f) framework must allow adjustment of the hardness of the problem instance to maintain the average block interval time,
        (g) framework must be scalable and 
        (h) all mining pools must solve tasks of equal hardness.''}
\end{quote}

\noindent In this work, we propose a framework \emph{ScaloWork}, which answers all these affirmatively. A comparative study with state-of-the-art is in Table~\ref{tab:my_label} enumerates the contributions of our work.


\subsection{Our Contributions} 
The MDS problem, as pointed out by~\cite{sym12111885,oliver2017proposal,wang2014domination,chinnasamy2019minimum,haynes2002domination,alipour2020distributed,chatterjee2023chrisimos} strikes an optimal balance between computational difficulty, verification ease, real-world utility, and adaptability to PoW systems. 
As discussed earlier, the prior work by~\cite{chatterjee2023chrisimos} explored designing a PoUW based on MDS problems, answering several critical questions that must be addressed before considering it as a replacement for hash-based PoW. 
Despite their work, unresolved challenges remain that need to be answered to advocate MDS problems as a viable alternative in practice. In this study, we take up those issues and answer all of them positively. 
We list our contributions below.

\begin{enumerate}
    \itemsep 0em
    \item[A.] We propose \emph{ScaloWork}, a framework for useful proof-of-work that is scalable and replaces hash-based protocol with the MDS problem on network instances. Any miner that submits the best solution within the stipulated time becomes the block proposer. To prevent solution stealing, our framework requires the utility company to submit isomorphic instances of the network so that each miner/mining pool gets a unique instance having equal hardness. It reduces the extra overhead introduced in \emph{Chrisimos}~\cite{chatterjee2023chrisimos}. 
    

    
    \item[B.] One major problem in conventional hash-based pool mining framework is the {\em free-rider} problem.
    A free-rider in a pool is a miner who claims a portion of the block reward by submitting partial proof-of-work that has no impact on winning the mining game. Our framework supports distributed pool mining, allowing individual miners to join different pools to collaboratively solve dominating set problems for large graphs, rather than mining independently.  
    We avoid the problem of free-rider as the task is distributed among all the miners in a pool, and each miner's contribution plays a role in winning the mining game. Any miner submitting a wrong result can be detected and eliminated. This has been discussed in details in Section~\ref{sec-free-rider} and~\ref{sec-phase-of-the-protocol}. The distributed pool mining enables \emph{ScaloWork} to find solutions for large, real-world networks quickly. This makes \emph{ScaloWork} a strong candidate for alternative mining.

    \item[C.] We provide a formal security analysis that shows our protocol is as secure as the hash-based PoW. Additionally, \emph{ScaloWork} allows utility companies to get the solution directly for their graph instance from the newly mined block. This was not possible in~\cite{chatterjee2023chrisimos} as the dominating set was fetched for a modified graph instance, and the utility company had to apply reverse engineering to get the solution for original instance.
    
    \item[D.] We also prove that the block addition time for \emph{ScaloWork} is signficantly less than that of \emph{Chrisimos}. 
    The storage overhead for \emph{Chrisimos} exceeds by $\mathcal{O}(K|V|)$ compared to \emph{ScaloWork}, where $K$ is the number of miners in the network and $|V|$ is the number of vertices in the graph instance. Experimental results support our claim. \textcolor{black}{Our code is available on GitHub\footnote{\url{https://github.com/subhramazumdar/Distributedpoolmining.git}}.}
\end{enumerate}

\section{Background and Notations}
\label{background}



\subsection{Graph Isomorphism}
A graph $G'(V',E')$ is said to be isomorphic to $G(V,E)$ if there is a function $f$ such that $\forall a,b \in V', (a,b) \in E'\Leftrightarrow  (f(a),f(b)) \in E$ \cite{fortin1996graph}. There is no known polynomial-time algorithm for graph isomorphism problems on general graphs, although that exists for some special graphs. The problem is in the NP class but not in the NP-Complete class. So the problem is considered to be in the NP-Intermediate complexity class. The best-known algorithm for graph isomorphism problem given by László Babai \cite{babai2016graph} has sub-exponential complexity of $2^{O((\log n)^c)}$. Helfgott \cite{helfgott2017isomorphismes} further claimed that $c=3$. 

 \subsection{Minimum Dominating Set Problem} 
 A dominating set $S_G$ of a graph $G(V, E)$ can be defined as a subset of vertices $V'$ such that, $V' \subseteq V$, and every vertex in $G$ is either in $V'$ or is adjacent to some vertex in $V'$ \cite{10.1007/978-3-540-30140-0_19}. 
 A set $S_G$ is a minimal dominating set of the graph $G$ if it does not contain any other dominating set as a proper subset. 
 A dominating set of $G$ of the lowest cardinality is called \emph{minimum dominating set} (MDS). 
 Computing an MDS is an NP-hard problem. The decision version of the problem, i.e., \emph{dominating set problem}, is NP-complete. It is defined as \emph{``Given a graph $G(V,E)$ and an integer $k$, does $G$ have a dominating set of size less than $k$?''}\cite{garey1979computers} Alon and Spencer \cite{alon2016probabilistic} state and prove an important result mentioned in the following theorem.
\begin{theorem}
\label{th1}
For a graph $G$ with $n$ vertices and minimum degree $\delta$ there exists a dominating set of size less than $k=\frac{n(1+\ln(1+\delta))}{1+\delta}$.    
\end{theorem} 

\subsection{Cryptographic Primitives}

\paragraph{Aggregate signature} BLS signature \cite{boneh2019bls} operates in a prime order bilinear pairing group and supports simple threshold signature generation, threshold key generation, and signature aggregation. 

Let $\mathbb{G}$ and $\mathbb{G}_T$ be a prime order group of order $q$. 
Then $e$ defined as $e:\mathbb{G} \times \mathbb{G} \rightarrow \mathbb{G}_T$ is called a degerate bilinear pairing, if for $g\in \mathbb{G}$, $e(g^a,g^b) = e(g,g)^{ab}$, and $e(g,g) = 1$ iff $g=1$.
Also, let $H: M \rightarrow \mathbb{G}$ be a hash function modelled as random oracle \cite{bellare1993random}.
Then a BLS signature on a message $\hat{m} \in M$ is defined as follows:
\begin{itemize}
    \item[] $\mathsf{KeyGen}()$: sample $sk \stackrel{\$}{\leftarrow} \mathbb{Z}_q$ and set $pk \leftarrow g^{sk} \in \mathbb{G}$.
    \item[] $\mathsf{Sign}$($\hat{m},sk$): $\sigma \leftarrow H(\hat{m})^{sk} \in \mathbb{G}$.
    \item[] $\mathsf{SigVrfy}$($\hat{m}',\sigma,pk$): \emph{accept} if $e(g_,\sigma) ~=~e(pk,H(\hat{m}'))$, else \emph{reject}.
\end{itemize}

\noindent\textbf{Signature aggregation} \cite{boneh2003aggregate}. Given triples $(pk_i, \hat{m}_i, \sigma_i)$ for $i= 1, \ldots, l$, anyone can aggregate the signatures $\sigma_1,\sigma_2,\ldots,\sigma_l$ into a short, convincing aggregate signature $\sigma$ by computing $\sigma \leftarrow \sigma_1 \sigma_2 \cdots\sigma_l \in \mathbb{G}$.

 For all $i \in [l]$, we verify $e(g,\sigma)\stackrel{?}{=}e(\Pi_{i=1}^l pk_i,H(\hat{m}))$, if $\hat{m}_i=\hat{m}$, else $e(g,\sigma)\stackrel{?}{=}\Pi_{i=1}^le( pk_i,H(\hat{m}_i))$, if messages are different. This scheme is secure against existential forgery with chosen message attacks if the Computational Diffie-Hellman or CDH problem is hard - 
\textcolor{black}{\emph{Given $g, h, g^\alpha \in \mathbb{G}$, it is hard to compute $h^{\alpha}$, where $\alpha \stackrel{\$}{\leftarrow} \mathbb{Z}_q$}.}


\subsection{Bitcoin Mining Pools}
A mining pool is a group of miners who work together to find the nonce to match the difficulty target in a PoW-based blockchain. 
On correctly solving the problem, the block rewards are then distributed among the miners. 
Pools were created when cryptocurrency mining reached a difficulty level that's almost impossible for a single miner to solve within the proposed time frame \cite{romiti2019deep}. 
This crowded small miners out of the competitive mining process, forcing them to work together to compete with the large mining firms.  
 
A mining pool is typically maintained and coordinated by a pool manager. 
The success of the pool relies on its computational power, requiring miners to commit their resources to solve the mining puzzle on behalf of the pool. 
To measure the computing power contributed by each miner, the pool manager introduces a simpler puzzle to solve and collects the solutions to this simpler puzzle, known as partial solutions. 
For example, the full solution to the mining puzzle might require finding a hash value less than 100. 
To gauge miners' activity, the pool manager may accept partial solutions with hash values less than 200, which is significantly easier to achieve. 
Miners submit these partial solutions, referred to as shares, to the pool manager. 
If one of these shares happens to be the full solution, the pool manager claims the reward and distributes it among the pool members based on their submitted shares and a predefined reward-sharing scheme.  

The design of the reward-sharing scheme is a critical component of any mining pool \cite{wang2019survey}. 
An effective scheme must ensure the financial sustainability of the pool while incentivising miners to dedicate their computational resources to honest mining for the pool. 
Mining pools use various reward schemes to distribute earnings among miners based on their contributions. 
Common schemes include \emph{Pay-Per-Share (PPS)}, which provides consistent payouts per share but at higher pool operator risk, and \emph{Proportional (PROP), where rewards are based on shares submitted during a round, though it’s prone to pool-hopping.} \emph{Pay-Per-Last-N-Shares (PPLNS)} rewards miners based on shares from the last $N$ submissions, encouraging loyalty and deterring pool-hopping \cite{schrijvers2017incentive}. Each scheme varies in complexity, payout predictability, and fairness, catering to different miner and pool operator priorities.


\subsection{Free-rider in Pool Mining}\label{sec-free-rider}

The free rider problem in Bitcoin mining pools stems from the simplicity and partial-verification nature of hash-based PoW \cite{fisch2017socially}. It results in inefficiencies, unfair rewards, and reduced network security. In mining pools, free riders refer to miners who participate in the pool to gain rewards without contributing their fair share of computational work \cite{johnson2014game}. These miners exploit the collective efforts of other participants while minimizing their own contribution, which disrupts the fairness of the system. A free-rider miner submits fake or low-effort shares (submissions showing partial progress towards solving the hash puzzle) to the pool. Hash-based PoW relies on miners solving a cryptographic puzzle and submitting ``shares'' of work to the mining pool. A ``share'' is a proof of partial work, typically a hash that meets a lower difficulty than the blockchain’s actual target. The issue arises because submitting shares does not necessarily equate to fair contribution. Free riders can submit fake shares (without performing the actual work), or shares generated by less expensive methods, such as reusing previous work. The current system also makes it harder to distinguish between genuine miners and free riders because both submit similar data structures (hashes).

Most pools distribute rewards based on shares rather than actual contributions to finding the block. This incentivizes miners to submit as many shares as possible, even if they are of low quality or require minimal effort. Free riders thus collect a proportional reward without significantly contributing to the block discovery. If free riders overrun mining pools, they may lose their ability to compete effectively with other pools \cite{eyal2015miner}. This increases the centralization of mining, making the network more vulnerable to attacks like 51\% attacks or censorship by dominant pools.
 \section{Formal Description of ScaloWork}
\label{solution}

\subsection{System Model and Assumptions}
There are three types of entities in our protocol: (a) a utility company supplying a graph instance, (b) an auditing committee selecting the graph instance, and (c) the miners of the Bitcoin network. 
\textcolor{black}{We assume the following about these entities.}
\begin{itemize}
\item The auditing committee must have less than $1/3$ Byzantine nodes to guarantee safety and liveness. Committee members run a Byzantine Fault Tolerant (BFT) consensus (e.g., Practical BFT) to select the next graph instance.
\item Finding the exact MDS is intractable for large graphs (standard NP-hardness assumption).
    \item Honest miners must control greater than $50\%$ of computational power and follow the protocol. Such miners aim to maximize rewards, not disrupt the network.
    \item Malicious miners behave arbitrarily, and collude with each other to attack the system.
    \item The communication model is partially synchronous. Messages arrive within a known time bound $\eta$, where $\eta>0$ is \textcolor{black}{known by all the participants of the protocol}. 

\end{itemize}
We describe the role of each entity in \emph{ScaloWork}.
\begin{itemize}
    \itemsep 0em
    \item {\em A utility company} could be any social networking company or company providing telecommunication services. We denote it as $U_{\mathcal{B}}$. Utility companies earn high profits by utilizing the dominating sets to realize their objective. Thus, the company shares a portion of its profit as remuneration with the miner who solves the dominating set for the given graph instance. Any utility company submits its graph instances to a common public platform along with a pseudonymous public key.

\item {\em An auditing committee}, $A_{\mathcal{B}}$, is elected to measure the hardness of the graph instance. The auditing committee members are sampled from the set of miners of the Bitcoin network. It is non-trivial to sample a committee with majority honest members in a permissionless setting. One cannot account for the network size, with miners dynamically entering or leaving the network. We use the idea followed in ByzCoin \cite{kogias2016enhancing}, where only the miners with dedicated resources (in this context, it is the mining power) can join the auditing committee. A time window is fixed within which a mining pool adding a block has a chance of being a member of the committee. The share proves the mining pool's membership in the group. Once a block is added, we shift the window by one block. Any pool possessing a share for an expired time window is not considered a part of the committee. Older shares expire as the window moves forward, ensuring that only recent and active mining pool participate in consensus. This limits the committee size and prevents inactive or historical mining pool from influencing decisions.
Once the committee is elected, they select a utility company for the current block proposal and check the hardness of the graph instance. Committee runs BFT consensus (e.g., PBFT) to seelect a graph based on the difficulty and hardness, sign and broadcast the chosen graph.

Given that at least $2/3$ members agree on the graph instance, adding trivial or stale graph instances would violate the safety property of the Bitcoin ledger. This will reduce the value of block rewards (as more blocks would be mined quickly), as miners will lose trust in the system. That will not be in the interest of strong miners who invest significant resources. The freshness of the committee members, reducing the risk of persistent malicious influence.


\item The third entity of this system is the miner that may join any {\em mining pool}. Joining a mining pool reduces the computational overhead on a miner as the task of finding a dominating set gets distributed among other pool members. Various mining pools of the Bitcoin network compete to perform the task of mining a block and once a block is propagated, it is validated by the miners.

\end{itemize}



\subsection{Phases of the Protocol}\label{sec-phase-of-the-protocol}
At a given moment, we can estimate the mining power of a miner based on the number
of blocks the miner has successfully mined within the
current window \cite{kogias2016enhancing}. Given that the collective hash power is relatively stable, each
active miner mines blocks statistically proportionate to the amount of hash power the miner
has contributed during this time window. We define the different phases of \emph{ScaloWork}:\\

(i) \emph{Preprocessing Phase}: This phases can be divided into following sub-phases:
\begin{itemize}
    \item \emph{Auditing Committee Selection}: The size $w$ of the share window is defined by the average block-mining rate over a given time frame. Once a block is added, the window is slid to accommodate the new block, and based on that, a new auditing committee is selected before proposing the next block. This mechanism limits the membership of the mining pool to recently active ones. We consider the size of the auditing committee $A_\mathcal{B}$ for proposing block $\mathcal{B}$ to be $c_m$. \textcolor{black}{Assuming that there are $K$ miners in expectation, we elect $c_m$ committee members so that at least $2/3$ members will be honest.} Each committee selects a utility company that has registered with the system. 
    \item \emph{$A_\mathcal{B}$ selects the utility company}: The auditing committee selects a valid utility company's public key. The selected utility company shares a tuple $P_G$, and signed hash of the tuple $P_G$ for the graph $G$, denoted as \textcolor{black}{$\sigma_{P_G}=Sign(H(P_G),sk_{U_{\mathcal{B}}})$}. $P_G$ comprises the reward for solving the problem $rd_G$, public key of utility company $pk_{U_{\mathcal{B}}}$, $n$ as the vertex count of $G$, number of isomorphic instances created for $G$ i.e., $z$, and other properties of $G$ like number of edges $m$, maximum and minimum degree, $\Delta$ and $\delta$ respectively. The committee members check the signature $\sigma_{P_G}$ and whether the graph $G$ submitted by $U_{\mathcal{B}}$ is sufficiently hard by setting a threshold for hardness. They may use the framework proposed in \cite{sym15010140} to estimate the difficulty, and any problem whose difficulty lies below the threshold is rejected. A metric to measure hardness is specified in the block header so that anyone can check if the instances are non-trivial. This could be in terms of the permissible range of the graph instance; for example, the graph instance must not have a vertex count less than 50000, and the average degree of the graph must not be less than 50 but not more than 150. \textcolor{black}{Other methods of analyzing the hardness of an instance may involve checking the degree distribution.}   

Once the auditing committee is convinced of the hardness of the instance, it generates a new identifier for the graph, $id$. It does an aggregate signature of all the members on $id\|H(P_G)$ denoted as $\sigma_{c_m}^G$. \textcolor{black}{This ensures that at least one honest member has signed.}
\item \emph{Instances for the block mining}: After the auditing committee has sent approval, the utility company dumps the graph instances $\langle (G_{1},\sigma_{G_1}),(G_{2},\sigma_{G_2}),\ldots, (G_{z},\sigma_{G_z}) \rangle$, into the problem pool, where each $G_i$ is isomorphic to the instance $G$ and $\sigma_{G_i}$ is the signature on $G_i$. $U_{\mathcal{B}}$ shares the storage address $addr_G$ from where the miner can fetch the instances. \textcolor{black}{$U_{\mathcal{B}}$ forms a transaction $\tau_{reward}$ with \emph{time-locked UTXO} that allows any miner to claim the reward who finds the best solution for a given instance after the timeperiod $T_{max}^G$ mentioned in the script. A lookup table is used to estimate $T_{max}^G$, which is discussed later. It is shared with the auditing committee $A_{\mathcal{B}}$. The script needs signature of all $|c_m|$ members of the auditing committee, denoted by public keys $pk_{m_1},pk_{m_2},\ldots,pk_{m_{|c_m|}}$, and also signed by the utility company. Any miner who gets the smallest dominating set in the given timelock claims the reward.}

\end{itemize}


(ii) \emph{Mining of Block}: A mining pool's manager $M$ receives $\tau_{reward}$ having an amount $rd_G$, and starts performing the following steps:
\begin{itemize}
    \item[(a)] $M$ checks if $\sigma_{P_G}$ and $\sigma_{c_m}^G$ are valid, parses $P_G$ to get information regarding $G$ and the address $addr_G$ from where it will fetch the graph instance. Additionally, $M$ checks if ${id}$ is greater than the instance ID of the previous block. 
\item[(b)] $M$ creates a block $\mathcal{B}$ with a set of transactions $\tau_G'$ comprising transactions from the mempool, and a puzzle-fee transaction $\tau_{rd,M}$. The transaction $\tau_{rd,M}$ spends the output of the reward transaction $\tau_{reward}$ to an address of the pool manager $M$. No two members of two different pool mining pool will have the same $\tau_{rd,M}$ even if the rest of the transaction selected from mempool remains the same.
\item[(c)] Once the block is formed, $M$ does a hash on the concatenation of the root of Merkle tree formed using $\tau_G'$ (denoted as $h_{\langle MR,\tau_G' \rangle }$) and hash of the previous block $\mathcal{B}_{prev}$ (denoted as $h_{\mathcal{B}_{prev}\textcolor{black}{)}}$, and then take a modulo over the number of instances in the problem pool. It gives the instance's index, say $j$, to be selected from $addr_G$. Once $M$ has fetched the instance $(G_j,\sigma_j)$ from $addr_G$, it checks if the specification of the size of the network, the minimum, and maximum degree of $G$ matches with that of $G_j$ and correctness of signature $\sigma_{G_j}$. It also checks if the size of $G_j$ is within the permissible range. 
\item[(d)] Once the pool manager is ensured of all the correctness, it distributes the task of finding a dominating set among the different miners in a pool. The miners in the pool tries to find the best solution for the given instance $G_{j}$ within a given block interval time $T_{max}^G$, as mentioned in the timelocked transaction $\tau_{reward}$. The criteria is that any solution fetched by the mining pool should be at most $\frac{n(1+\ln(1+\delta))}{1+\delta}$. 

Distribution of task in the pool depends on the chosen algorithm for finding the solution. Each miner is responsible for making decisions about addition of few nodes in the dominating set. The correctness of the solution depends on miners sharing updates about their nodes and neighborhood states. 


\item[(e)] Once the solution is computed, it is communicated back to $M$, who checks the correctness of the solution and adds the solution for the dominating set of $G_j$, the bound on the size of the dominating set, the characteristics of the graph specified in $P_G$, the hash of the previous block $h_{\mathcal{B}_{prev}}$, Merkle root of the transaction set $\tau_G'$ denoted as $h_{\langle MR, \tau_G' \rangle}$, $addr_G,id,\sigma_{P_G},$ and $\sigma_{c_m}^G$ in the block header of $\mathcal{B}$ and broadcasts it to the rest of the network. 
\end{itemize}
The pseudocode is defined in Algorithm ~\ref{gen_block}. The output of the dominating set for graph $G$ is $S_G$. This is a function definition for block mining that any mining pool must follow. We do not provide a function definition for FindDominatingSet since each mining pool decides the algorithm it would select for solving the graph instance.

\paragraph{Free-rider Resistant Pool Mining} 
\textcolor{black}{A distributed greedy algorithm for finding a dominating set in the context of a useful Proof-of-Work system can eliminate the free rider problem in a mining pool by designing the mining process in a way that requires each miner to actively contribute to the computation. The distributed greedy algorithm for finding a dominating set works by dividing the vertex set of the graph into smaller subsets and assigning them to individual miners in a pool. Miners collaborate to collectively compute the dominating set of the entire graph. The correctness of the solution depends on miners sharing updates about their nodes and neighborhood states. Lack of contribution from free riders will lead to incorrect solutions. From the lack of information on the subset of vertices assigned to a free rider, it will be easy to point out the free riders and penalize them. The problem scales well because each miner handles only a small subset of graph vertices. As the pool size increases, the workload is distributed more evenly. Each miner contributes a specific, measurable part of the solution. Rewards can be distributed proportionally based on miners' contributions.}

\RestyleAlgo{ruled}
\begin{algorithm}[!ht]
\SetKwRepeat{Do}{do}{while}
			\caption{{\sf Block Generation}}
			\label{gen_block}
   \textbf{Input}: Transaction $\tau_{reward}$, public keys of $|c_m|$ members $pk_{m_1},pk_{m_2},\ldots,pk_{m_{|c_m|}}$ of the committee, public key $pk_{U_{\mathcal{B}}}$ of utility company, reward $rd_G, P_G$, $addr_G$, $id$, $\sigma_{P_G}$, $\sigma_{c_m}^G$ from the transaction.
   \begin{itemize}
       \item \textcolor{black}{Parse $P_G$ to get public key of utility company $pk_{U_{\mathcal{B}}}$, $n$ as the vertex count of $G$, number of isomorphic instances created for $G$ i.e., $z$, and other properties of $G$ like number of edges $m$, maximum and minimum degree, $\Delta$ and $\delta$ respectively}
       \item \textcolor{black}{start\_time=current\_timestemp}
   \end{itemize}

		 \If{ $\mathsf{SigVrfy}(H(P_G),\sigma_{P_G},pk_{U_{\mathcal{B}}}) = 1$ and $\mathsf{SigVrfy}(id\|H(P_G),\sigma_{c_m}^G,\langle pk_{m_1},\ldots,pk_{m_{|c_m|}} \rangle) = 1$ }
                {
                \begin{itemize}
                    \item \textcolor{black}{Form a set of transaction $\tau_G'$ by selecting transaction from mempool and including block reward $rd_G$} 
                 \item \textcolor{black}{Compute $h_{\langle MR,\tau_G' \rangle }=MerkleRoot(\tau_G')$}
               
                \item \textcolor{black}{Get index $j=H(h_{\langle MR,\tau_G' \rangle }\|h_{\mathcal{B}_{prev}})\mod z$}
                 \item \textcolor{black}{Fetch the instances $G_j(V_j,E_j)$ from $addr_G$}
                 
                 \item \textcolor{black}{Set $S_G=|V_j|$ and compute $k=\frac{n(1+\ln(1+\delta)}{1+\delta}$}
                \end{itemize}
                 
                 \While{$ |S_G|> k$ and time elapsed $<T_{max}^G$}{
                    $S_G'\leftarrow \textrm{FindDominatingSet}(G,pool)$\\
                     \If{$|S_G'| \leq k$}
                     {
                     
                         $S_G\leftarrow S_G'$\\
                  }
                    
                       end\_time=current\_timestemp\\
                
                }
                 \If{end\_time-start\_time $<T_{max}^G$ and $ S_G \leq k$}
                 {
                 \begin{itemize}
                     \item \textcolor{black}{$\textrm{block\_header}=HeaderGen(h_{\textrm{prev}\_\mathcal{B}},h_{\langle MR,\tau_G' \rangle },S_G,P_G)$}
                     \item \textcolor{black}{$\mathcal{B}\leftarrow BlockGen(\textrm{block\_header},\tau_G')$}
                 \end{itemize}
                     \Return{$\mathcal{B}$}
            }
            \Else
            {
               abort
          }
          }
	\end{algorithm}

\RestyleAlgo{ruled}
\begin{algorithm}[!ht]
			\caption{{\sf Block Verify}}
			\label{verify}
   \textbf{Input}: $\mathcal{B},past\_size_{DS}$ 
   \begin{itemize}
       \item \textcolor{black}{Parse $\mathcal{B}$ to get $h_{\mathcal{B}_{prev{}}},h_{\langle MR,\tau_G' \rangle }$, $\tau_G'$, public keys of $|c_m|$ members $pk_{m_1},pk_{m_2},\ldots,pk_{m_{|c_m|}}$ of the committee, $pk_{U_{\mathcal{B}}}$ of utility company, reward $rd_G, P_G$, $addr_G$, $id$, $\sigma_{P_G}$, $\sigma_{c_m}^G$, $S_G$ and $T_{max}^G$}
   \item \textcolor{black}{Parse $P_G$ to get public key of utility company $pk_{U_{\mathcal{B}}}$, $n$ as the vertex count of $G$, number of isomorphic instances created for $G$ i.e., $z$, and other properties of $G$ like number of edges $m$, maximum and minimum degree, $\Delta$ and $\delta$ respectively}
   \item \textcolor{black}{Set $visited\_set=\phi$}    
   \end{itemize}
			
  \If{ $h_{\langle MR,\tau_G' \rangle } \neq MerkleRoot(\tau_G')$ or current time $\geq T_{max}^G$ or ${id}\leq GetPrevBlockGraphid()$ or $past\_size_{DS}<|S_G|$ or $\mathsf{SigVrfy}(H(P_G),\sigma_{P_G},pk_{U_{\mathcal{B}}})\neq 1$ or $\mathsf{SigVrfy}(id\|H(P_G),\sigma_{c_m}^G,\langle pk_{m_1},pk_{m_2},\ldots,pk_{m_t} \rangle)\neq 1$ or current\_timestemp$>T_{max}^G$}
                {
                    reject solution                    
                }
                \Else
                {\begin{itemize}
                    \item \textcolor{black}{Get index $j=H(h_{\langle MR,\tau_G' \rangle }\|h_{\mathcal{B}_{prev}})\mod z$}
                \item  \textcolor{black}{Fetch the instances $(G_j,\sigma_{G_j})$ from $addr_G$}
                
                \end{itemize}
                 \If{$\mathsf{SigVrfy}(H(G_j),\sigma_{G_j},pk_{U_{\mathcal{B}}})\neq 1$}
                 {
                            reject solution
                 }
                \For{$v$ in $S_G$}
                {
               \begin{itemize}
                    \item \textcolor{black}{Mark $v$ as visited} 
                  
                         \item \textcolor{black}{Add $v$ to $visited\_set$}
                \end{itemize}
                    
                    \For{$v' \in N(v)$}
                    {
                    \begin{itemize}
                        \item \textcolor{black}{If $v'$ is not visited then mark it visited }
                         \item \textcolor{black}{Add $v'$ to $visited\_set$}
                    \end{itemize}
                         
                    }

                }
                
                    \If{$|visited\_set|=|V|$}
                    {
                            
                            $past\_size_{DS}=|S_G|$\\

                    }
                    \Else
                    {
                        reject solution
                    }
                    
                    }

	\end{algorithm}

(iii) \emph{Block Verification}: The pseudocode is defined in Algorithm~\ref{verify}. The verifier performs the following steps:
\begin{itemize}
    \item[(a)] The verifier initializes the variable $past\_size_{DS}=\infty$. Upon receiving the first block, it checks if the transactions in the block, $\tau_G'$ are valid and constructs the Merkle tree using $\tau_G'$.
    \item[(b)] From the block header, the verifier gets the dominating set of a graph isomorphic to $G$, $id$, $h_{\langle MR, \tau_G' \rangle}$, $addr_G,id,\sigma_{P_G},$ and $\sigma_{c_m}^G$. 
    \item[(c)] It derives the index $G_{id}$ from $H(h_{\mathcal{B}_{prev}}\|h_{\langle MR, \tau_G' \rangle}) \mod z$, download the graph $G_{id}$ from address $addr_G$, checks if ${id}$ is greater than the instance ID of the block added previously.
    \item[(d)] The verifier matches the hardness of the graph instance as specified in the block header and the specification defined in $P_G$. If any of the checks fail, the block is marked invalid. It checks whether the block has a valid dominating set and is within the bound specified in the header. The cardinality of this dominating set is assigned to $past\_size_{DS}$. Now, the verifier will cache this block until it gets a block with a dominating set of size smaller than $past\_size_{DS}$.
    \item[(e)] The verifier will continue to check for new blocks till the $T_{max}^G$ expires. After this, it will accept the last block it had stored. Any solution appearing after $T_{max}^G$ is rejected. Miners who verify the block reach a consensus on the best result in the given epoch, and a new block is added to the blockchain.

\end{itemize}

\paragraph{Block Reward} The block reward in a distributed greedy algorithm system for finding dominating sets can be designed to fairly compensate miners, and prevent exploitation by free riding. The pool manager can define a reward-distribution mechanism based on the existing reward policies. If some miners are assigned a higher number of nodes, the manager can allocate higher rewards to balance fairness. A mining pool that submits a dominating set having the least cardinality within $T^G_{max}$ wins the mining game and gets the block reward. This incentivizes mining pools to compete with each other to find the best solution. The block reward comprises a fee from the transaction set in the block and the fee provided by the utility company for solving the dominating set of the graph. The lookup table provides an estimate of the block interval time for a graph instance, indirectly providing some insight into the hardness of the problem. We expect a fair utility company to decide on the remuneration directly proportional to the hardness of the problem, and it should be more than the current reward offered by coinbase transactions. 


Mining pools may collude and send out random solutions without doing any work. However, even if a single pool behaves honeslty then it is enough to foil the entire
colluding effort. Colluding mining pools are not likely to cooperate and do the DoS attack, especially if they know they only need a relatively small amount of useful work to win the competition and claim the rewards. The rationale behind the behavior is to start competing and return the best solution. Thus, the mining game induces competition among the mining pools, and the framework acts like a \emph{decentralized and distributed minimal dominating set solver}.  

\subsection{Constructing the Lookup Table}

We record the block interval time for each benchmark instance in the lookup table as it was done in \emph{Chrisimos} \cite{chatterjee2023chrisimos}. The block interval time estimates the time for (a) block generation and (b) block verification. 


      
      


(a) \emph{Block Generation}: An exhaustive search for finding a dominating set in $G(V,E)$ of size $\frac{n(1+\ln(1+\delta))}{1+\delta}$ takes exponential time in the size of the input. Thus, the miners can use any algorithm that returns a result in polynomial time. We apply the distributed greedy heuristic \cite{lynch2004distributed} to estimate the runtime for several synthetically generated datasets to find a minimal dominating set for $G$. This may not be the minimum since the greedy algorithm does not guarantee an optimal result.

We start with an empty set $S_G$, and then we greedily add ``good'' nodes to \( S_G \) until it becomes a dominating set. Nodes in \( S_G \) are referred to as \textit{black}, nodes covered by \( S_G \) (i.e., neighbors of nodes in \( S_G \)) are called \textit{grey}, and all uncovered nodes are \textit{white}. The span of a node \( v \), denoted by \( w(v) \), is defined as the number of white nodes among \( v \)'s direct neighbors, including \( v \) itself. So, the span of a non-black vertex is a non-negative integer, but the span of a black vertex is always $0$.  The span of a node can only decrease if any nodes within a distance of at most 2 are added to the dominating set. Therefore, if the span of \( v \) is greater than the span of any other node within a distance of at most 2, the greedy algorithm prioritizes \( v \) over its neighbors. This idea leads to a simple distributed version of the greedy algorithm, where each node \( v \) executes the following steps.


\RestyleAlgo{ruled}
\begin{algorithm}
\caption{Distributed Greedy Algorithm}
\begin{algorithmic}[1]
\STATE $S_G=\phi$
\WHILE{$\exists\ v \in V \ | \ w(v)>0$}
    \STATE \textbf{Calculate} $w(v)$ and \textbf{Send} to $N(v)\cup N(N(v))$
    \IF{$w(v) \geq w(v') \ \forall \ v' \in N(v)\cup N(N(v))$}
    \STATE $S_G=S_G\cup \{v\}$
    \STATE Color all \textit{white} neighbors of $v$ with \textit{grey}
    \ENDIF
\ENDWHILE
\end{algorithmic}
\end{algorithm}
The greedy strategy ensures that nodes with the highest span are prioritized, which quickly covers a large number of white nodes in each step. The use of hop-2 neighbors prevents conflicts where nearby nodes might try to join the dominating set simultaneously.
Breaking of ties by unique IDs ensures fairness and avoids deadlocks or repeated conflicts.

\begin{theorem}
The Distributed Greedy Algorithm computes a \( \ln \Delta \)-approximation for the minimum dominating set problem in \( O(n) \) rounds.
\end{theorem}
The approximation factor of \( \ln \Delta \) comes from the greedy strategy's efficiency in reducing the number of uncovered nodes. This is the same bound achieved in the sequential greedy algorithm. Since the distributed version replicates the greedy decisions, it inherits this approximation factor. Each node performs a constant number of local computations (e.g., computing its span and comparing it with others). Since there are $n$ nodes, in the worst case, the process requires $O(n)$ rounds, as the algorithm terminates when all white nodes are covered. This happens after at most $n$ iterations, one for each node potentially joining the dominating set.

 (b) \emph{Block Verification}: Any miner receiving the block checks if the dominating set returned by (a) is valid.

 \textcolor{black}{Given the block generation and verification on a benchmark instance $G$ with a runtime $\tau$ units, and the upper bound on size of the dominating set is $k$, a mining pool will try to find the dominating set of size at $k'\leq k$. To ensure that the some mining pool will definitely return a dominating set of cardinality less than the upper bound, we propose to set the block interval time $T_{max}^G$ to $l\tau$ where  $l \in \mathbb{R}^{+}, l>1$. This will ensure that the estimated block interval time for a new graph instance is sufficient for a pool to find an acceptable solution.} 

\textbf{Estimating Block Interval Time}
\textcolor{black}{It is necessary to estimate the block interval time to determine the feasibility of the solution.} Once the lookup table is constructed, we use the same to estimate the block interval time for any new graph instance. In \emph{ScaloWork}, each mining pool is assigned a different isomorph of the input graph, and each miner in the pool works cooperatively to find a good dominating set. Given an instance $G''(V'',E'')$ to the network, pool manager searches for the entry $G'(V',E')$ in the lookup table with the maximum vertex count less or equal to $|V''|$ and extracts the block interval time $T_{max}^{G''}$ for $G''$. The proposed time for a new block corresponding to $G''(V'',E'')$ will be $T_{max}^{G'}*\frac{|E''|\times |V''|}{|E'|\times |V'|}$. 
\textcolor{black}{An estimation of the block interval time assures a mining pool of the time bound provided to fetch a solution without unnecessarily wasting the computation resource. }

\begin{lemma}
\label{x}
    For a graph instance $G$, the block time interval $T_{max}^G$ is sufficient for adding the block to the Blockchain.
\end{lemma}
\begin{pro}
During the estimation of block interval time, we showed that it considers block generation and verification time. If the block generation time is $\tau$ (if the graph instance is already present in the lookup table), we set $T_{max}^G$ to $l\tau: l>1$. If the graph instance is absent in the lookup table, the time is estimated by scaling it based on the edge count and vertex count of the graph instance. Since the lookup table is prepared using a greedy heuristic, a rational miner will get a solution by at least using the greedy heuristic. 
\end{pro}

\begin{lemma}
\label{cf1}
The dominating set of a graph $G$ is within the bound stated in ~\cref{th1} for ScaloWork, whereas the dominating set of $G$ was within the bound in expectation for Chrisimos.   
\end{lemma}
\begin{pro}
    \emph{Chrisimos} requires the miners to find a minimum dominating set for the extended graph with vertex count twice that of the original graph. After a miner has won the mining game and added a block, the utility company has to extract the solution for the original graph.

    A guarantee on the bound of the solution in expectation does not ensure that the bound will hold for all the cases. It is possible that the dominating set for extended graph is not a good solution for the original graph. This problem does not exist in \emph{ScaloWork} as the miner solves the problem on an isomorphism of the original graph instance. If the utility company retains the mapping, it can retrieve the solution. 
\end{pro}
\begin{lemma}
Ratio of time taken for the block generation in ScaloWork and Chrisimos is approximately $\frac{1}{\delta_{avg}+\frac{\delta}{2}}$. 
\end{lemma}
\begin{pro}
    The extended graph of $G(V,E)$ in \emph{Chrisimos}, denoted as $G_T(V_T,E_T)$ where $|V_T|=2|V|$ and $|E_T|=2|E|+\delta \frac{|V|-1}{2}$ \cite{chatterjee2023chrisimos}. The extend function in \emph{Chrisimos} takes $\mathcal{O}(|E_T|)$ and hence dominates the run time of block generation. We can replace $|E|=\delta_{avg} |V|$. On the other hand, the greedy heuristic for finding a dominating set in graph $G$ takes $\mathcal{O}(|V|)$. This is the asymptotic runtime for block generation in \emph{ScaloWork}. If we take the ratio of the run time, we have the expression $\frac{|V|}{2|E|+\delta \frac{|V|-1}{2}}= \frac{|V|}{\delta_{avg}|V|+\delta \frac{|V|-1}{2}}$. Assuming $|V|-1\approx |V|$, the ratio is $\frac{1}{\delta_{avg}+\frac{\delta}{2}}$.
\end{pro}

\begin{lemma}
 \textcolor{black}{Ratio of time taken for the block verification in ScaloWork and Chrisimos asymptotically tends to $\frac{1}{2}$}.
 \end{lemma}
 \begin{pro}
      The verifier needs to check the neighbors of the vertices in the dominating set. The time complexity is the summation of the degrees of vertices in dominating set, without any double counting of a vertex. \textcolor{black}{We transform the original graph but do not increase the size in \emph{ScaloWork}. In \emph{Chrisimos}, the graph is transformed by adding additional vertices and edges, and the transformed graph has twice the vertex count compared to the original graph. Hence, the ratio asymptotically tends to $\frac{1}{2}$.}
 \end{pro}


\section{Security Analysis}
\label{sec}
Security of \emph{ScaloWork} relies on the assumption that the underlying chains operate under a well-designed incentive mechanism that ensures an honest majority of miners and no mining pool holds more than 50\% of computation power, guaranteeing consistency and liveness. Security also depends on the hardness of graph isomorphism, which prevents a mining pool from stealing another's solution when a new block propagates in the network. We consider that a Bitcoin network has about ten to fifteen mining pools that run the vast majority of the network. Note that each of those pools usually consists of thousands of individual miners from across the world. The exact number of individual computers contributing to the network according to a recent estimate is around 70,000 that runs Bitcoin mining software \cite{bitpanda_mining_pools}. \textcolor{black}{Also, a graph of size $n$ will have $n! = \mathcal{O}(n^n)$ possible isomorphism. The graph instance provided by the utility company will have orders of more than $1000$, so the number of possible isomorphisms exceeds the number of mining pools.} 

\subsection{Security in Terms of Solution Integrity}
 \begin{theorem}
 \label{th-miner}
\textcolor{black}{Two mining pools will never receive the same graph instance.}
\end{theorem}
 \begin{pro}
  \textcolor{black}{Two mining pools will get the same graph instances only when they get the same ID upon hashing the concatenation of the previous block's hash in the chain and root of Merkle tree built from the transaction set in the block.
  \textcolor{black}{Now the transaction tree contains the unique coinbase transaction for every pool, which makes the Merkle root unique.}
  Any standard root of the Merkle tree has a size of 256 bits. The hash function SHA-256 is collision-resistant, and part of the preimages uses the same previous block hash. Even if both the mining pool use the same transaction set, they will have different coinbase transactions. So, they will have different preimages with same prefix. Hence, no two mining pool will get the same ID.}   
 \end{pro}

\begin{theorem}
\label{th22}
    \textcolor{black}{Given that a mining pool knows that two instances, $G$ and $G'$, are scale-free and isomorphic, the pool can find the mapping between the two graphs within the block interval time $T_{max}^G$ with negligible probability}.
\end{theorem}
\begin{pro}
\textcolor{black}{The number of vertices in $G$ and $G'$ are $n$, so the search space for finding the mapping between $G$ and $G'$ is $n!$ which is $\mathcal{O}(n^n)$. Even if $n=100$, the execution time is more than $2^{600}$. On the other hand, $T_{max}^G$ must have a feasible limit (for Bitcoin, this is 10 mins on average), keeping the property of liveness in mind. Any mining pool can't try all the mapping within $T_{max}^G$. It may try out other possible heuristics, such as determining the highly connected vertices in both graphs. Even if the pool figures out $k$ such highly connected vertices where $k<<\frac{n}{2}$, this will reduce $n$ to $n-k$, but still, search space remains exponential. The best-known algorithm for graph isomorphism problem \cite{babai2016graph} has sub-exponential complexity of $2^{O((\log n)^c)}$, where $c\geq 1$ and later it was shown $c=3$ \cite{helfgott2017isomorphismes}. If we consider $n=1000$, the run time will still be of the order of $2^{1000}$. Even though a quasi-polynomial algorithm is faster than an exponential algorithm, the runtime is still larger than any polynomial time algorithm. Thus, a pool will succeed in finding the mapping between the two graphs within $T_{max}^G$ with negligible probability}. 
\end{pro}


\begin{theorem}
\label{iso1}
\textcolor{black}{Given two isomorphic and scale-free networks $G$ and $G'$, and the dominating set of $G$ is revealed to the mining pool $M$, then the probability for the pool to infer the dominating set of $G'$ from the given information within $T_{max}^G$ is negligible.}
\end{theorem}
\begin{pro}
In a scale-free network, few vertices have a large degree compared to others. This is due to the nature of the graph where few nodes are highly trusted, and new nodes tend to join highly trusted nodes, inducing the \emph{rich becomes richer} phenomenon. Let us assume that a pool can figure out $b$ such highly connected nodes that are part of the dominating set for the graph $G$. If all of them have distinct degrees, then the mapping is straightforward. But if any $b'$ out of $b$ nodes have the same degree, then the pool has $b'!$ possibilities, i.e., $\mathcal{O}(b'^{b'})$. Given that other nodes in $G$ and $G'$ can have other not-so-highly connected nodes with degrees ranging between $\delta$ and $\Delta$, there would be many possible mapping combinations. If the number of such vertices $b'=\alpha n$ where $\alpha<1$, there would be $(\alpha n)!$ possibilities, which leads to exponential search. So, even if the instances are scale-free, the probability of finding the mapping within time $T_{max}^G$ is negligible.
\end{pro}

\subsection{Security in Terms of Safety and Liveness}
We define the security goals of our protocol:\\
(i) \emph{Safety}: Honest miners do not commit different blocks at the same height.\\
(ii) \emph{Liveness}: If all honest miners in the system attempt to include a certain input block, then, after a few rounds, all miners report the input block as stable.

We prove the property of safety and liveness in the synchronous model. We leave this analysis in the asynchronous setting as a part of the future work.

\paragraph{Chain selection rule in \emph{ScaloWork}}
\label{chain}
Two or more pools may solve their instances at about the same time and publish their blocks, creating the situation known as a fork in blockchain systems. Forks are usually resolved in the synchronization phase using the rule specified for the particular Blockchain. Only one of the blocks will pass both the verification and synchronization phases. We define a chain selection rule whereby the maximum work done chain is selected. 
Work done in a chain is the summation of the work done in the individual blocks forming the chain. The work done in a block is proportional to the graph's size and the pool's effort in finding a dominating set of lower cardinality. Thus, we define the work done for the block $B$ as $WD_B=|E| \times |V| \times \frac{\frac{n(1+\ln(1+\delta))}{1+\delta}}{|S_G|}$.
  
If there is a fork at block $B'$, then the pool chooses the chain, starting from $B'$, having the highest work done, i.e. if $\mathcal{C}=\{C_1,C_2,\ldots,C_m\}$ be $m$ such forks from block $B'$, then choose the chain $C_i \in \mathcal{C}$ such that $\sum_{B \in C_i: C_i \in \mathcal{C}} WD_B$ is maximum. An honest pool considers a block $B$ committed if $B$ is buried at least $f$ blocks deep in its adopted chain. We assume $f$ to be at least six block confirmations to guarantee security as good as hash-based PoW.

We summarize the rules of a valid chain as follows:\\
 (i) The Verifier rejects a block with a graph instance with an ID either less than or equal to the graph ID of the instance mined in the previous block of the main chain. The verifier rejects the block if the graph instance has a malformed signature (not signed by the committee). \\
 (ii) \emph{A block is said to be confirmed if it has received at least $f$ confirmations}: We assume all honest miners reach an eventual finality over a single chain up to a block that has $f$ confirmations.\\
  (iii) \emph{If the miners observe another sub-chain with higher work done}:  If any block in the other sub-chain violates rule (i), discard the chain, else miners consider the other chain as the highest work done.

\textcolor{black}{\emph{Resolving Forks}: If two distinct pools find the dominating set of least cardinality in the given time interval simultaneously, then it might lead to a fork in the Blockchain network. It depends on how well a block propagates across the network. If there is a fork, it will eventually be resolved with the chain selection rule proposed here. Eventually, one chain will have the highest work done, and a mining pool will start mining on this chain. Blocks in the discarded chain become orphan blocks \cite{neudecker2019short}.}  

\paragraph{Safety Property}
Since all the graph instances are solved and added sequentially into the blockchain, we prove the safety property in selfish-mining attacks \cite{nicolas2019comprehensive}. \textcolor{black}{ By selfish mining, strong nodes can gain
higher payoffs by withholding blocks they create and selectively postponing their publication.}

\begin{lemma}
\label{sf}
 Given that the signatures of the committee members are unforgeable and no mining pool holds more than \textcolor{black}{50\% computation power in the blockchain network}, the probability of a selfish mining attack is negligible.
\end{lemma}
\begin{pro}
 Suppose the malicious miner (or pool) induces a fork from the parent block $\mathcal{B}_{prev}$. The chain of the adversary remains the same till block $B_{prev}$ and starts differing from here, so we label these blocks as $B_{prev+i}', 1\leq i\leq z$ where the length of the private chain. Given $ z\geq L$ where $L$ is the length of the existing longest sub-chain $\langle B_1,B_2,\ldots,B_L \rangle$, as each graph instance is provided sequentially after elapse of block interval time. For the subchain $\langle B_{prev+1}',B_{prev+2}',\ldots,B_{prev+z}' \rangle$ to be selected, $\sum_{i=1}^z WD_{B_{prev+i}'}>\sum_{i=1}^L WD_{B_{i}}$. \\
 
 If the adversary finds that the cumulative work done in his private sub-chain is less than the cumulative work done in the main chain, it is highly likely he will abandon selfish mining and try to add the new block on the main chain. To continue the attack, the adversary has to match the work done of the existing longest subchain, and that would require the miner to re-mine the block where the difference occurred. Suppose at block $k<z$, $\sum_{i=1}^k WD_{B_{prev+i}'}<\sum_{i=1}^k WD_{B_{i}}$, then adversary has to find the solution for block $B_{prev+k}'$ that is as good as the solution for $B_{k}$. If the adversary has a graph instance whose index does not match with the index of the instance in block $B_{k}$, then it either has to find a better solution or find the mapping of vertices of instance in $B_{k}$ to its instance in $B_{prev+k}'$. From ~\cref{iso1}, we know that the adversary would need to compute exponential possibilities in the worst case to find out the mapping. 
 
 Additionally, a miner can't generate several graph instances and mine a longer chain. It follows from the rule (i) of \emph{chain selection rule}, where a verifier will reject any illegitimate graph instance not signed by the committee members. Since we assume that the majority of the committee members are honest and a secure signature scheme is used to sign the instance, miners will not be able to forge signatures for all the instances. Hence, the adversary can pull off the attack with negligible probability.
\end{pro}
\textcolor{black}{Building a secret chain in an attempt to reverse payments is called a double
spending attack. A malicious party successfully creates an alternative (and dishonest) version of the blockchain ledger, which includes the malicious/dishonest transaction which \emph{double-spends} the coins in another transaction on the existing longest chain \cite{sapirshtein2017optimal}. It would be difficult to tell which version of the blockchain ledger is the ``correct'' one. In \emph{ScaloWork} we choose the chain with the most accumulated work as the ``correct'' one. With the most-accumulated-work rule, at least the payee can choose to wait for $f$ more confirmations to make his risks lower.} If a pool attempts selfish mining, other honest miners (pools) can continue solving the problem collaboratively, invalidating the selfish miner's advantage. The selfish mining pool risks missing out on rewards because its private solution may not align with the final, validated solution from the network. This contrasts with hash-based PoW, where a valid block is independent and immediately verifiable.

\subsection{Security Against Collusion Attacks}
 If a utility company acts like a Bitcoin network miner, it has the advantage of having prior knowledge of the graph. However, the utility company provides the fee for mining the block. So, the company does not get any additional fee for winning the mining game, except for the transaction fee. Given that the company already has obtained the \emph{best solution} for a given graph instance in one of the mined blocks, it may tend to reuse the graph instance in some other block. This is termed as \emph{replay attack}. The success of the attack is again subjected to the fact that the epoch when the utility company decides to put profitable transactions in the block must match the selection by the auditing committee. If the utility company is elected at a stage when all the profitable transactions have already been mined, then it has to search for suitable unspent transactions from the mempool. However, if any other mining pool manages to supply a better solution within the given epoch, then the company will lose the mining game.
    
    \textcolor{black}{The utility company might be interested in biasing the game if it is interested in earning a substantial fee by packing a certain number of high-valued transactions in its block. The success depends on the fact that no miner would be able to win the mining game by submitting a better solution. Other factors influencing the participation of the utility company in this biased game will be the expected gain and also the probability of getting re-elected by the committee.} 
\begin{claim}
\label{c1}
  Given \textcolor{black}{that more than $2/3$ members of auditing committee are honest, a malicious utility company (also a miner in the network), upon re-election,} will perform a ``replay attack'' if and only if the transaction fee offered for the block is greater than the reward offered for solving the graph instance for the given block.
\end{claim}
\begin{pro}
     Given that the malicious utility company (who is also a miner in the network) got re-elected by the auditing committee and has supplied a graph instance for which it had obtained a solution in one of the previous blocks $B'$, $\lambda$ is the computation power of the pool that had mined the block $B'$. This pool has a chance of again winning the mining game. Thus, with probability $(1-\lambda)$, the utility company may lose the mining game in this block interval time.
     $F$ is the transaction fee by mining all transactions in the current block formed by utility company, and $rd_G$ is the reward offered for the graph instance $G$ in the given block. If the utility company wins the mining game, it implies that no miner has been able to find a solution better than the one provided in block $B'$. If there exists a pool who found a solution better than the one existing in block $B'$, the utility company loses the mining game. We calculate the expected payoff of the utility company as $\lambda F - (1-\lambda) rd_G>0$. The attack is possible if $F> \frac{1-\lambda}{\lambda} rd_G$. Given that $rd_G$ is comparable to the Bitcoins minted upon mining a block in hash-based PoW and $\lambda_{max}=0.5$, the least value of $\frac{1-\lambda}{\lambda}$ tends to 1, so the transaction fee of a block has to be large enough to incentivize the utility company from mounting this attack.
     
\end{pro}

\section{Performance Analysis}
\label{performace}

\begin{figure*}[!ht]

    \centering
    \begin{subfigure}[b]{0.31\textwidth}
    \centering
    {\includegraphics[height=1.2in]{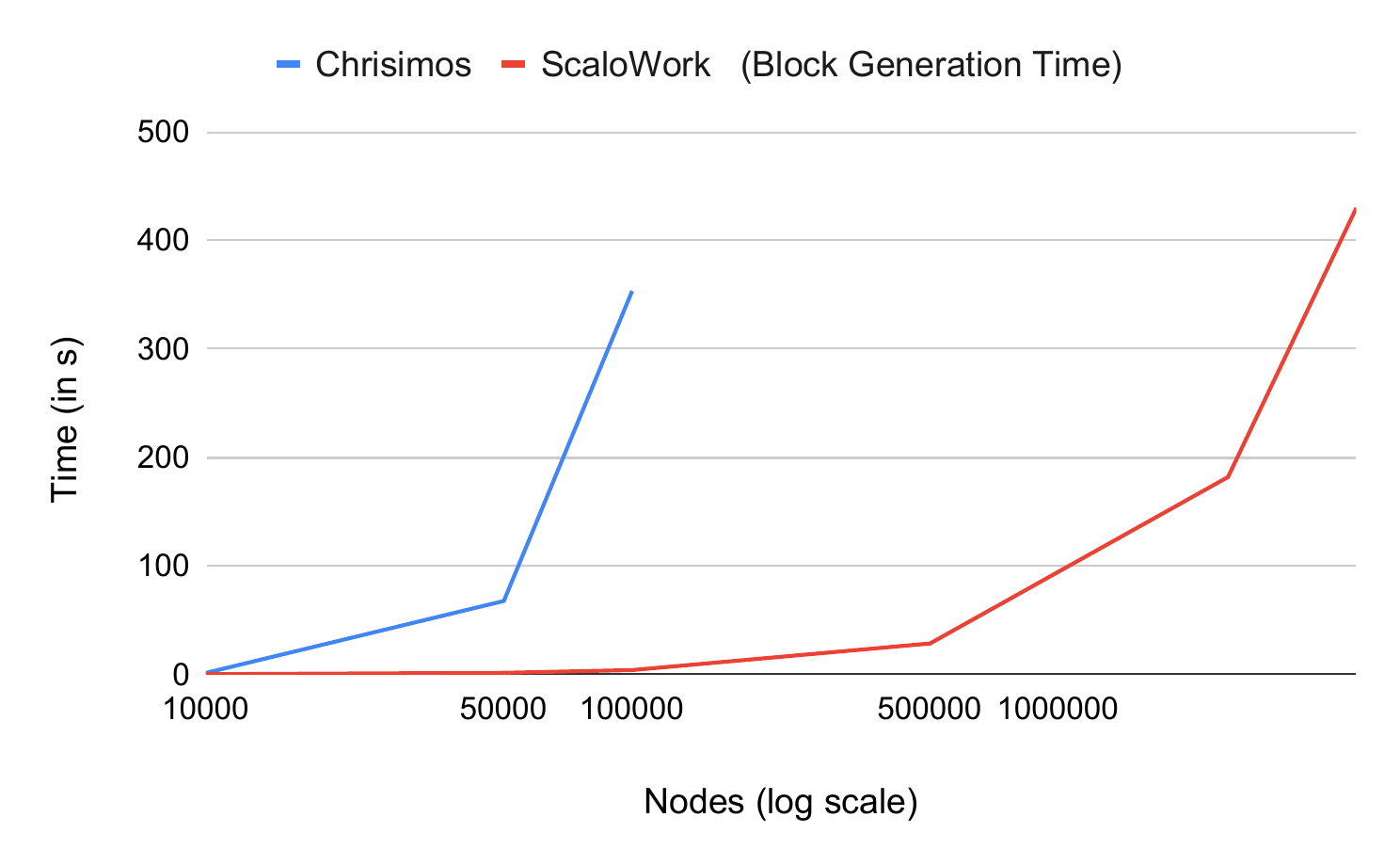}  }
    {\includegraphics[height=1.2in]{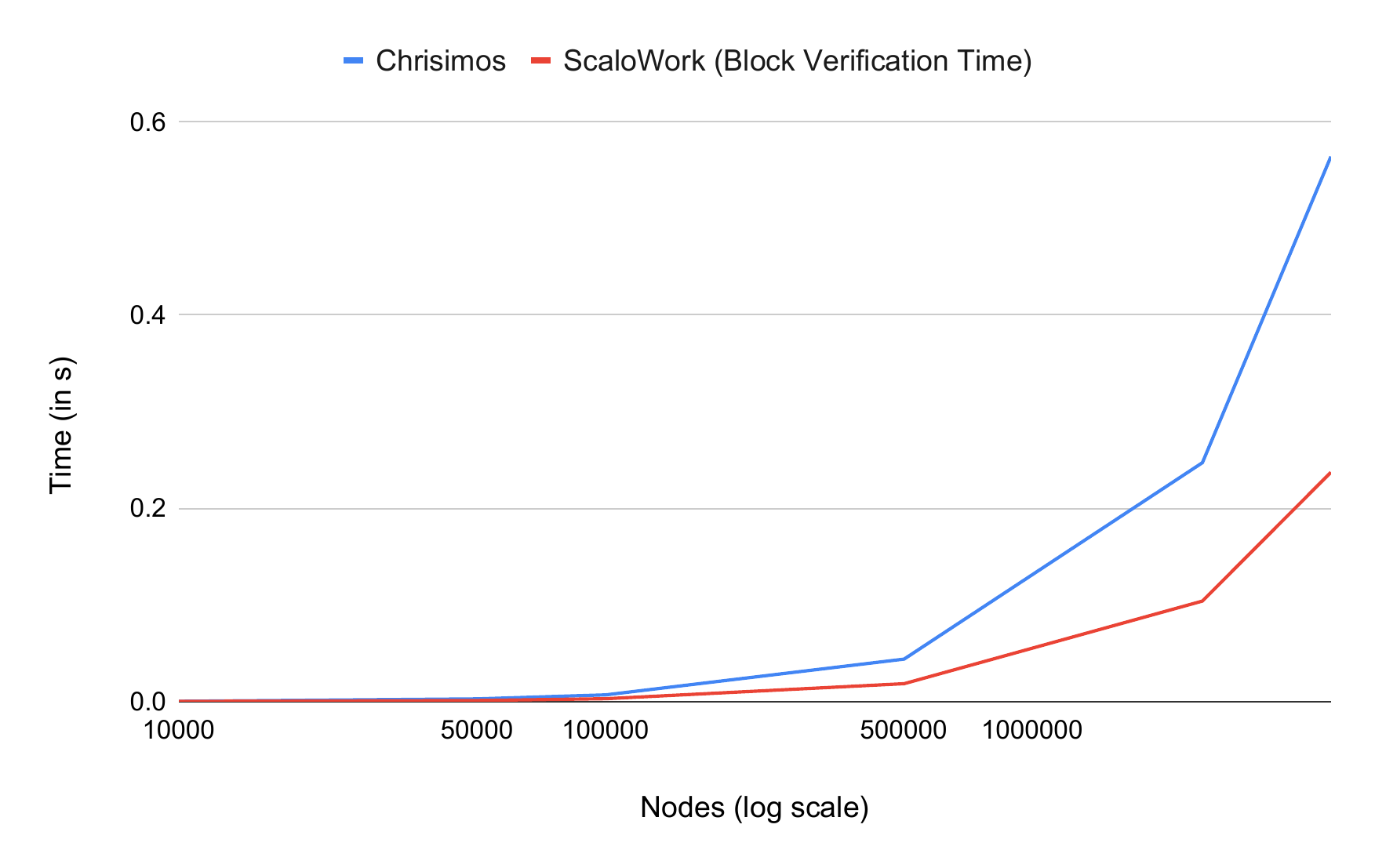} }   
    \label{march20}    
     \caption{Average degree 50}
\end{subfigure}
    \begin{subfigure}[b]{0.31\textwidth}
    \centering
    {\includegraphics[height=1.2in]{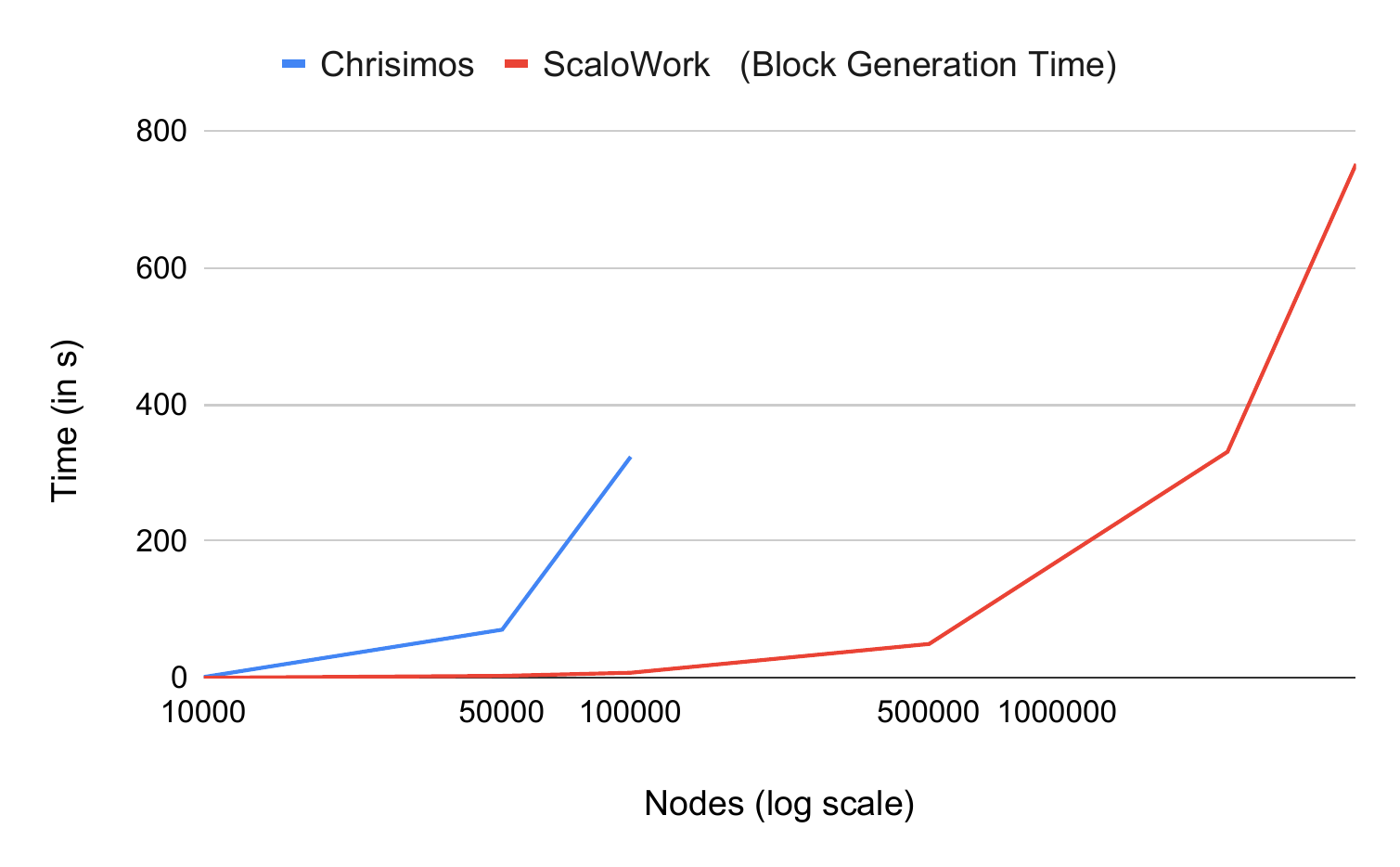} }
    {\includegraphics[height=1.2in]{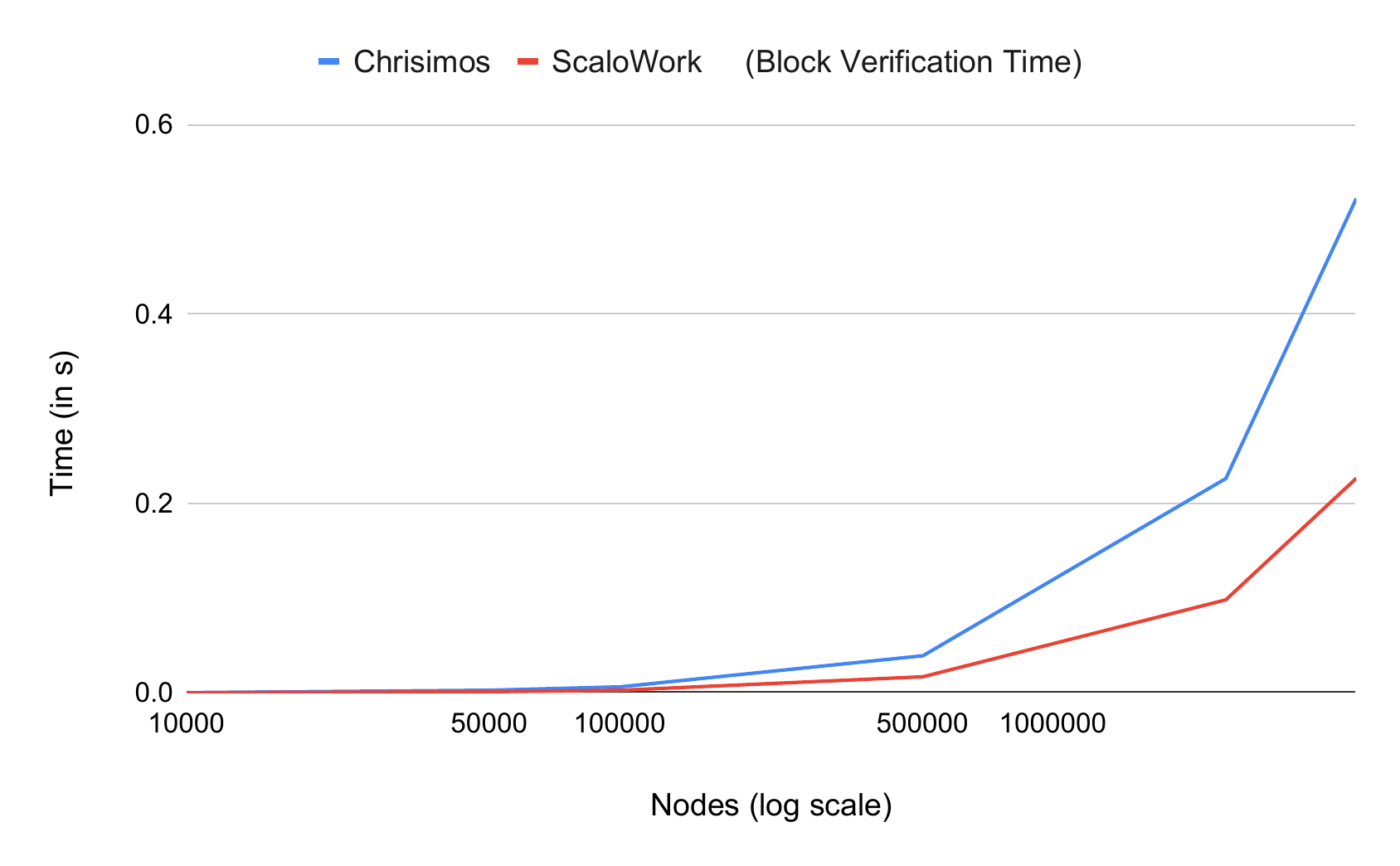}  }
    \label{april21}
     \caption{Average degree 75}
\end{subfigure}
\begin{subfigure}[b]{0.31\textwidth}
    \centering
{    \includegraphics[height=1.2in]{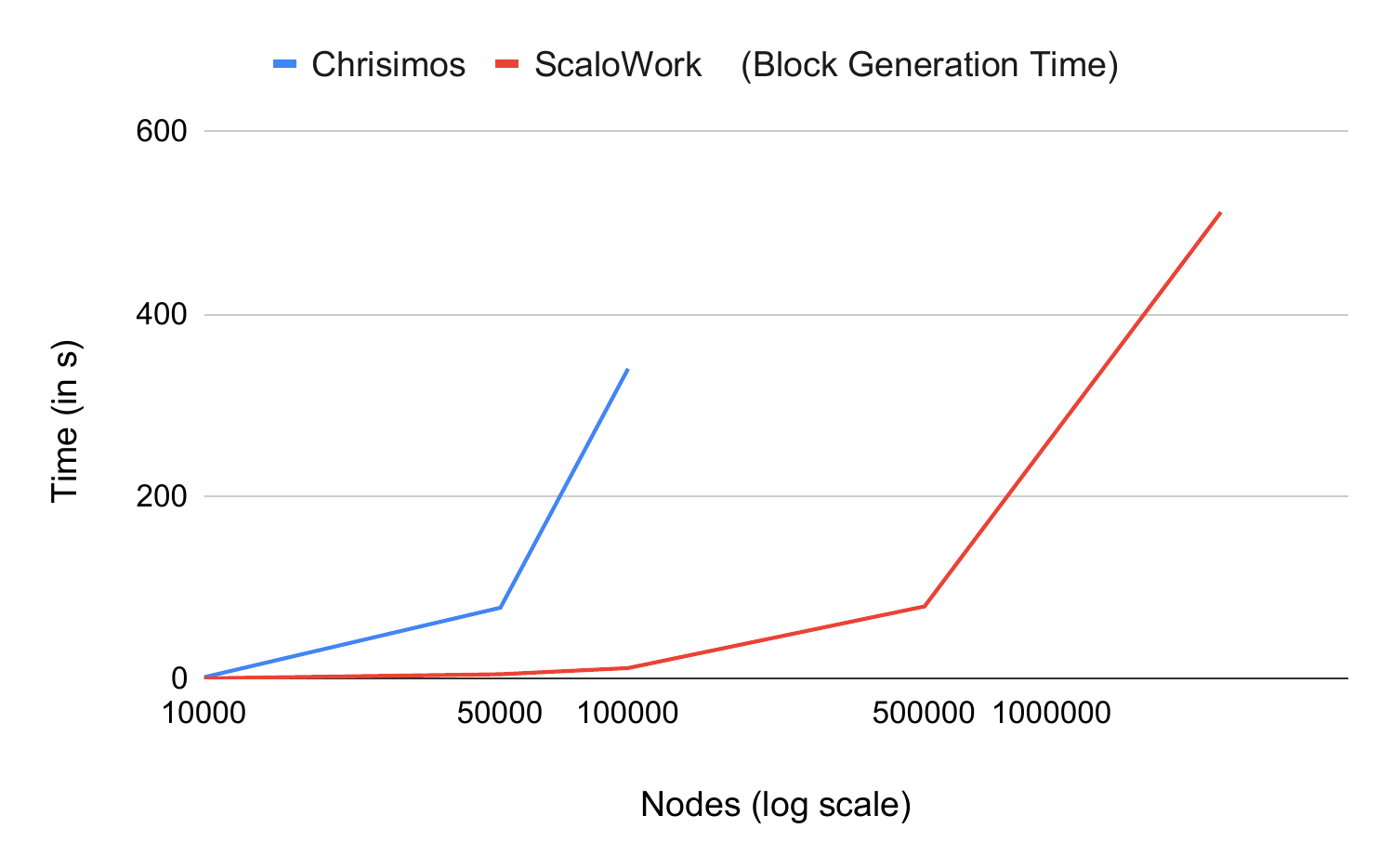}  }
{    \includegraphics[height=1.2in]{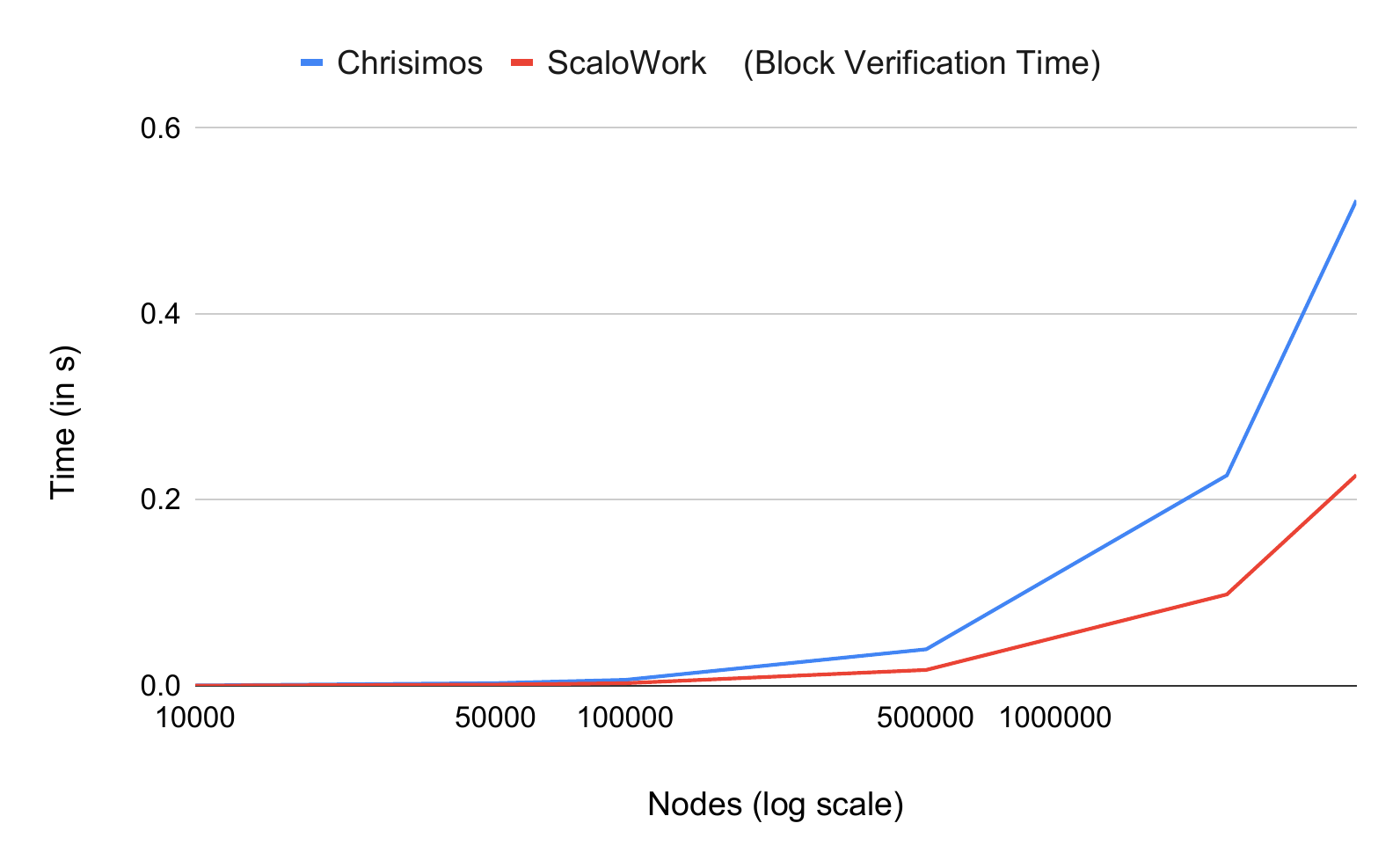}}

     \caption{Average degree 100}
\end{subfigure}
\caption{\textcolor{black}{Comparing block generation time and verification time for Chrisimos and ScaloWork}. We set the cut-off for run time to 15 mins. Thus, the block generation time for \emph{Chrisimos} in \ref{simulated}(a), (b), and (c) stops at node count 100000.}
\label{simulated}    
\end{figure*}

(a) \emph{Setup}: For our experiments, we use igraph \footnote{\url{https://igraph.org/}} and OpenSSL \footnote{\url{https://openssl.org/}} for network analysis and mathematical operations. The system uses processor AMD Ryzen Threadripper PRO 7965WX 24-Cores. We use OpenMP \footnote{\url{https://www.openmp.org/}} for computation of the distributed greedy algorithm for dominating set. This package simplifies the implementation of parallelism using 48 threads. Social networks and other utility networks follow the power-law model, where few vertices are central to the graph instance. Thus, we use synthetically generated graph instances (based on Barab{\'a}si-Albert Model \cite{albert2002statistical} and Erdős-Rényi Model \cite{seshadhri2012community}) mimicking this model to generate the benchmark datasets. We choose appropriate parameters to generate the synthetic graph instances such that they simulate real-life networks. The number of nodes varies between 100000 and, 5000000 (X-axis in log-scale), and the average degree of the graph varies between 50 and 100. We ignore the time taken in the preprocessing phase. \textcolor{black}{We assume that the graph instances are selected by the committee way before the epoch for a new block starts}. Once a new block is added, the auditing committee will not change significantly. \textcolor{black}{Note that the selection of a specific set of parameters does not imply that our framework will not work for other parameters, one can vary the number of nodes as well as the average degree. Our code is publicly available \footnote{\url{https://github.com/subhramazumdar/Distributedpoolmining.git}}.} 

(b) \emph{Observations}: We compare our scheme \emph{ScaloWork} with \emph{Chrisimos}. ~\cref{simulated}(a) shows the block generation (top figure) and verification time (bottom figure) when average degree is 50, ~\cref{simulated}(b) shows the block generation (top figure) and verification time (bottom figure) when average degree is 75, and ~\cref{simulated}(c) shows the block generation (top figure) and verification time (bottom figure) when average degree is 100. The cutoff time for our experiment is set to 15 mins, so any time after 15 mins is not reported. For average degrees 50, 75, and 100, \emph{Chrisimos} exceeds the cutoff time of 15 mins at 500000 nodes (reporting around 197 mins), highlighting scalability challenges. \emph{ScaloWork} consistently handles larger graph sizes, scaling up to 5000000 nodes without exceeding the cutoff time (around 12 mins) but till average degree 75. However, \emph{ScaloWork} exceeds this limit when average degree is 100. We also observe that \emph{Chrisimos} struggles with higher-degree graphs during block generation. \emph{ScaloWork} shows robustness across different average degrees, indicating that the distributed pool mining approach effectively handles the added complexity introduced by higher degrees. The block verification time for \emph{ScaloWork} is less than half of the time taken by \emph{Chrisimos}. Block verification time is negligible compared to generation time, being below 1s for both \emph{Chrisimos} and \emph{ScaloWork}.

(c) \emph{Discussions}: The block generation in \emph{Chrisimos} depends on solo mining, with a miner exploring all the edges in the
given graph instance. Extending the graph results in a doubling of vertex count. So if we
report the result for vertex count 100000, it denotes the execution time of finding and verifying the dominating set on a graph of size 200000. \emph{ScaloWork} does not require an extension on the original graph, and the workload is distributed across miners in a mining pool. We consider 48 threads acting as different miners in a pool, involved in running the algorithm for finding the dominating set. This drastically reduces the time taken for block generation. The verification time increases slowly (but linearly) compared to the block generation time with an increase in the size of the input graph. The reason is that the verifier checks whether the vertices in the dominating set cover the entire
graph. The time complexity is bounded by the number of vertices in the graph, which is less than the number of edges. 
The overall performance analysis highlights the advantages of \emph{ScaloWork} in terms of scalability, efficiency, and real-world applicability, making it a more viable framework compared to \emph{Chrisimos} for practical PoW-based blockchain systems.

\textcolor{black}{\emph{Storage overhead:} In \emph{Chrisimos}, the extended graph instance has at least $2|E|+\frac{\delta(|V|-1)}{2}$ edges, so each miner must have a storage capacity to store the instance. If the network has $K$ miners and utility company stores just the original graph having $|E|$ edges, then total storage capacity needed is at least $K(2|E|+\frac{\delta(|V|-1)}{2})+|E|$.
 In the worst case, a mining pool will comprise one miner, and in such a situation \emph{ScaloWork} requires the miner to have storage capacity just as large as the original graph. Utility company stores $K$ isomorphisms of the original graph. Total storage capacity is $2K|E|$ edges. Thus, \emph{Chrisimos} demands an extra capacity of at least $K\frac{\delta(|V|-1)}{2}$ edges.}

\section{Conclusion}
\label{conclusion}

In this paper, we present \emph{ScaloWork}, a novel framework for Useful Proof-of-Work  that transforms blockchain mining into a meaningful computational process by centering it around the dominating set problem. Our framework addresses critical limitations of traditional hash-based PoW by ensuring that computational efforts contribute to solving real-world problems while maintaining the security, decentralization, and fairness essential for blockchain systems. By leveraging graph isomorphism, we provide miners with unique problem instances, ensuring solution extractability and protecting miners’ contributions during block propagation. The distributed pool mining approach enables scalability, allowing collaborative problem-solving for large, real-life graphs and effectively addressing free-riding issues. Our design ensures equitable reward distribution, aligning closely with existing Bitcoin pool mining practices while enhancing fairness and efficiency.  

We demonstrate through formal security analysis and experimental evaluations that \emph{ScaloWork} is as secure as hash-based PoW while significantly improving computational efficiency and scalability. Compared to existing frameworks like \emph{Chrisimos}, \emph{ScaloWork} offers superior performance, storage efficiency, and adaptability to real-world applications. Looking forward, we believe the principles underlying \emph{ScaloWork} can be extended to other NP-complete problems in graph theory, such as the clique problem or graph coloring. As part of our future work, we aim to explore our framework for solving such problems.
 

\section*{Acknowledgment}
We thank Department of Computer Science and Information Systems, BITS Pilani, KK Birla Goa Campus, Goa, India
for funding our research.


\end{document}